\documentclass[twocolumn,showpacs,superscriptaddress,prb]{revtex4-1}
\usepackage{graphicx}
\usepackage{subcaption}
\usepackage{color}
\usepackage{amsmath,amsthm,amssymb}
\usepackage{braket}
\usepackage[colorlinks,citecolor=blue,linkcolor=red]{hyperref}

\usepackage{booktabs}
\usepackage{makecell,multirow}

\usepackage{float}

\graphicspath{ {images/} }

\usepackage{soul}
\begin{document}
 
\title{Fully-Compensated Ferrimagnetic Spin Filter Materials within the Cr\textit{M}\textit{N}Al Equiatomic Quaternary Heusler Alloys }

\author{Gavin Winter}
\thanks{These two authors contributed equally}
\email{winter.ga@northeastern.edu}
\affiliation{Department of Physics, Northeastern University, Boston, MA 02115, USA}

\author{Matthew Matzelle}
\thanks{These two authors contributed equally}
\email{matzelle.m@northeastern.edu}
\affiliation{Department of Physics, Northeastern University, Boston, MA 02115, USA}

\author{Christopher Lane}
\email{laneca@lanl.gov}
\affiliation{Theoretical Division, Los Alamos National Laboratory, Los Alamos, New Mexico 87545, USA}
\affiliation{Center for Integrated Nanotechnologies, Los Alamos National Laboratory, Los Alamos, New Mexico 87545, USA}

\author{Arun~Bansil}
\email{ar.bansil@neu.edu}
\affiliation{Department of Physics, Northeastern University, Boston, MA 02115, USA}

\date{version of \today} 
\begin{abstract}
XX'YZ equiatomic quaternary Heusler alloys (EQHA's) containing Cr, Al, and select Group IVB elements (\textit{M} = Ti, Zr, Hf) and Group VB elements (\textit{N} = V, Nb, Ta) were studied using state-of-the-art density functional theory to determine their effectiveness in spintronic applications. Each alloy is classified based on their spin-dependent electronic structure as a half-metal, a spin gapless semiconductor, or a spin filter material. We predict several new fully-compensated ferrimagnetic spin filter materials with small electronic gaps and large exchange splitting allowing for robust spin polarization with small resistance. CrVZrAl, CrVHfAl, CrTiNbAl, and CrTiTaAl are identified as particularly robust spin filter candidates with an exchange splitting of $\sim 0.20$ eV. In particular, CrTiNbAl and CrTiTaAl have exceptionally small band gaps of $\sim 0.10$ eV. Moreover, in these compounds, a spin asymmetric electronic band gap is maintained in 2 of 3 possible atomic arrangements they can take, making the electronic properties less susceptible to random site disorder. In addition, hydrostatic stress is applied to a subset of the studied compounds in order to determine the stability and tunability of the various electronic phases. Specifically, we find the CrAlV\textit{M} subfamily of compounds to be exceptionally sensitive to hydrostatic stress, yielding transitions between all spin-dependent electronic phases. 
\end{abstract}


\maketitle 
\section{Introduction}\label{sec:intro}

Equiatomic quaternary Heusler alloys (EQHA's) provide a relatively unexplored rich phase space which may harbor novel electronic, magnetic, topological, and correlated phases of matter. Recently, EQHA's have attracted attention due to their variety of low energy spin-polarized transport and tunneling properties. This makes them ideal candidate materials for the next generation of spintronic and quantum information processing applications,\cite{Bainsla,Xu,Gao} including direct use in magnetoresistive random access memory (MRAM), spin diodes and transistors, and magnetic tunnel junction (MTJ) devices.\cite{Hirohata}

MTJ's are the standard generator of spin-polarized currents and therefore the fundamental building blocks of MRAM. They are fabricated by inserting a thin barrier in between two electrodes in order to suppress the transmission rate for electrons of one spin over the other. Therefore, when a potential difference is applied between the electrodes a spin polarized current flows. Over the years, developments in functional configurations of magnetic and electronic properties for the electrodes and the barrier\cite{Worldege,Mooderaeus,Mooderarev} have pushed spin-polarization very close to 100\% at varying temperatures.\cite{Miao} However, room temperature MTJ's without net magnetism and high polarization are still elusive.

Fully-compensated ferrimagnets,\cite{Galanakis2015} exhibiting an asymmetric gap between the two spin channels, provide a clear route to discriminate between electronic spin states by yielding a highly asymmetric spin-dependent tunneling rate. Since the spin polarization is generated by intrinsic exchange splitting with no net moment, this class of materials can be used in nano scale devices without the need to correct for the deleterious effects of stray coercive fields. Since MTJ's are one of the most critical components for spintronic applications, the accurate prediction of new candidate barrier materials is of paramount importance. 

Many half-metals,\cite{deGroot} spin gapless semiconductors (SGS),\cite{Wang2008} and spin filter materials (SFM)\cite{Galanakis2016} have been predicted, but only a few have been synthesized.\cite{Stephen2016,Stephen2019,Stephen2019PRB,Tsuchiya}  
CrVTiAl, a member of the XX'YZ EQHA family, has been of particular interest due to its prediction to be a fully-compensated ferrimagnetic SFM with exceptional properties.\cite{Galanakis2015,Stephen2016,Cao} However, when synthesized in bulk form only a small component of an ordered L2\textsubscript{1} or Y phase was present. Similarly, the thin film was found in an SGS phase due to atomic swapping between magnetic sub-lattices,\cite{Stephen2019PRB} illustrating the key role the arrangement of the atomic species across the four magnetic sub-lattices plays in the electronic and magnetic properties of the quaternary Heusler alloys. Therefore, further study of the Cr-based EQHA's is needed to find other atomic combinations which are immune to disorder effects.

In this article, we predict 27 new spin-gapless semiconductors, half metals, and spin-filter material candidates within the equiatomic quaternary Heusler alloy family. Specifically, the electronic and magnetic properties of Cr$MN$Al compounds with $M$ and $N$ selected from Group IVB and Group VB elements, respectively, is presented. Most compounds were predicted to be fully-compensated ferrimagnets, following the variant of the Slater-Pauling rule\cite{Galanakis2013}  for $18$ valence electrons, with only four materials displaying a finite net magnetic moment. In particular, CrVZrAl, CrVHfAl, CrTiNbAl, and CrTiTaAl are identified as robust spin filter candidates, the latter two exhibiting $\sim 0.10$ eV band gaps, and all four with an exchange splitting of $\sim 0.20$ eV. Additionally, their properties are maintained in two of the three atomic arrangements, making them partially immune to atomic site disorder. Finally, we find the CrAlV\textit{M} subfamily of compounds to be exceptionally sensitive to hydrostatic stress, yielding transitions between all spin-dependent electronic phases. 

The outline of this paper is as follows. In Sec.~\ref{sec:compdetails} the computational details are summarized. In Sec.~\ref{sec:groundstate} the crystal structure is introduced along with a comparison of the various atomic orderings. The ground state electronic and magnetic structure for each compound studied is broken down into classes and presented in subsections A-C. Section~\ref{sec:stress} delineates the effects of hydrostatic stress on the CrAlV\textit{M} subfamily of compounds. Finally, Sec.~\ref{sec:conclusion} is devoted to the conclusions.

\begin{figure}
\includegraphics[scale=.5]{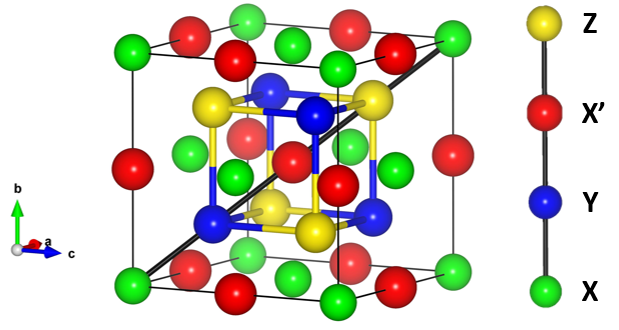}
\caption{(color online) Crystal structure of the equiatomic quaternary Heusler alloy in spacegroup $\text{F}\bar{4}3m$ (216) with Wyckoff positions $a$, $b$, $c$, and $d$ denoted by the green, red, blue and yellow spheres. The [111] body diagonal is indicated by the thick black line within the unit cell. The ordering of elements XYX'Z is shown on the right.}
\label{fig:struct}
\end{figure}

\section{Computational Details}\label{sec:compdetails}

{\it Ab initio} calculations were carried out by using the pseudopotential projector-augmented wave method\cite{Kresse1999} implemented in the Vienna ab initio simulation package (VASP) \cite{Kresse1996,Kresse1993} with an energy cutoff of $400$ eV for the plane-wave basis set. Exchange-correlation effects were treated using the strongly constrained and appropriately normed (SCAN) meta-GGA scheme.\cite{Sun2015} An 18 $\times$ 18 $\times$ 18 $\Gamma$-centered k-point mesh was used to sample the Brillouin zone. All atomic sites in the unit cell along with the cell dimensions were relaxed using a conjugate gradient algorithm to minimize the energy with an atomic force tolerance of 0.008 eV/{\AA}. When the energy difference between subsequent electronic steps was less than 1 $\mu$eV self-consistency was determined to be achieved. Pulay stress in VASP was used to simulate hydrostatic stress on the lattice, with relaxations being performed after each increase in stress. The Wigner-Seitz magnetic moment is the total moment of the unit cell, as defined by the Bravais lattice vectors, while the magnetic moments on atoms are the moments on spheres defined by the Wigner-Seitz radius in the pseudopotential. Crystal structure images were rendered using VESTA.\cite{vesta}

\section{Electronic Structure and Magnetic Ordering in Ground State}\label{sec:groundstate}

\begin{figure}[b]
\includegraphics[scale=0.3]{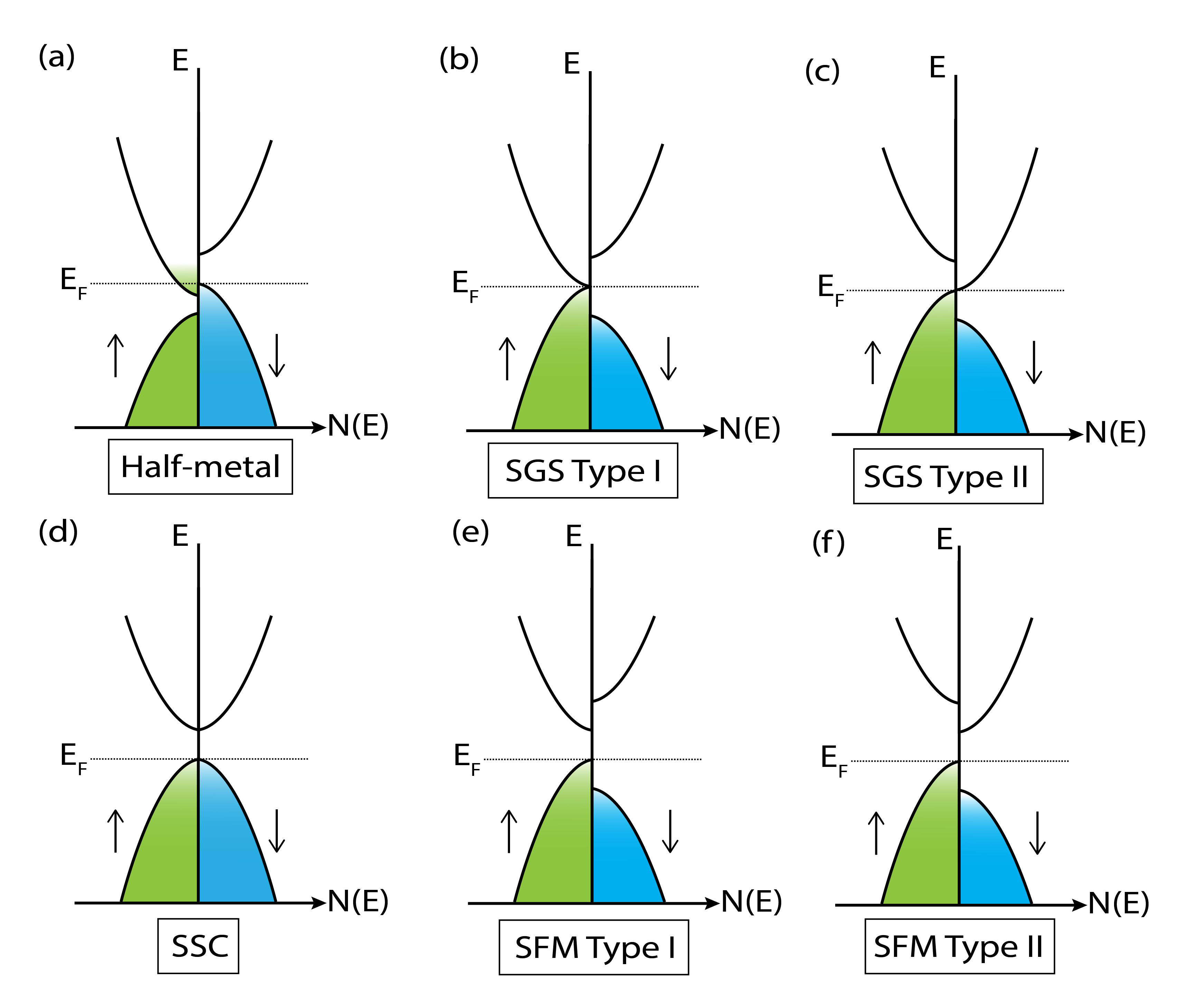}
\caption{(color online) Schematic of the various electronic phases: (a) half-metal, (b) spin gapless semiconductor (SGS) Type I, (c) spin gapless semiconductor (SGS) Type II, (d) symmetric semiconductor (SSC), (e) spin filter (SFM) Type I, (f) spin filter (SFM) Type II. Note that classification as a half-metal does not necessitate the presence of a gap below the Fermi energy.}
\label{fig:phase}
\end{figure}

\begin{table*}[htbp]
	\begin{center}
		\caption{Theoretically predicted electronic and magnetic properties of EQHA's of the three prototypical arrangements:  Cr\textit{N}\textit{M}Al, Cr\textit{M}\textit{N}Al, and CrAl\textit{N}\textit{M} (\textit{M} = Ti, Zr, Hf and \textit{N} = V, Nb, Ta).}
		\label{tab:table1}
		\scriptsize
		\begin{tabular}{l|c|c|ccc|c|cccc|c}
			%\hline
			%\toprule % <-- Toprule here
			\specialrule{1pt}{0.5pt}{0.5pt}
			\textbf{Arrangement} & \makecell{\textbf{Lattice} \\ \textbf{parameter} \\ \textbf{({\AA})}} & \makecell{\textbf{Electronic} \\ \textbf{phase}} & \multicolumn{3}{c|}{\textbf{Band gap (eV)}} & \makecell{\textbf{Exchange splitting} \\ \textbf{$\mathbf{2 \Delta E_{ex}}$ (eV)}} & \multicolumn{4}{c|}{\makecell{\textbf{Magnetic Moments} \\ \textbf{($\mathbf{\mu_{B}}$)}}} & \makecell{\textbf{Wigner-Seitz} \\ \textbf{Magnetic Moment} \\ \textbf{($\mathbf{\mu_{B}}$)}} \\
			
			\textbf{XX'YZ} & & & \textbf{$\uparrow$} & \textbf{$\downarrow$} & \textbf{Total} & & & & & \\
			
			\hline
			%\midrule
			CrVTiAl & 6.191 & SFM II & 0.730 & 0.966 & 0.558 & 0.430 & -3.305 & 2.384 & 0.539 & 0.005 & 0 \\
			CrVZrAl & 6.392 & SFM I & 1.055 & 1.508 & 1.055 & 0.237 & -3.395 & 2.584 & 0.333 & 0.008 & 0 \\
			CrVHfAl & 6.329 & SFM I & 1.078 & 1.337 & 1.078 & 0.216 & -3.293 & 2.539 & 0.259 & 0.009 & 0 \\
			\hline
			%\midrule
			CrNbTiAl & 6.372 & SFM I & 0.267 & 0.362 & 0.248 & 0.114 & -3.173 & 1.433 & 1.039 & 0.012 & 0 \\
			CrNbZrAl & 6.559 & SFM II & 0.610 & 0.757 & 0.443 & 0.166 & -3.203 & 1.702 & 0.633 & 0.023 & 0 \\
			CrNbHfAl & 6.495 & SFM II & 0.584 & 0.685 & 0.403 & 0.181 & -3.060 & 1.665 & 0.497 & 0.025 & 0 \\
			\hline
			%\midrule
			CrTaTiAl & 6.333 & SFM II & 0.190 & 0.266 & 0.152 & 0.038 & -3.013 & 1.264 & 1.073 & 0.016 & 0 \\
			CrTaZrAl & 6.516 & SFM II & 0.456 & 0.749 & 0.261 & 0.195 & -3.022 & 1.513 & 0.669 & 0.028 & 0 \\
			CrTaHfAl & 6.456 & SFM II & 0.361 & 0.621 & 0.120 & 0.240 & -2.869 & 1.483 & 0.525 & 0.029 & 0 \\
			\specialrule{1pt}{0.5pt}{0.5pt}
			%\hline
			%\bottomrule
			
			%\hline
			%\midrule
			CrTiVAl & 6.159 & SGS I & 0 & 0.577 & 0 & 0.342 & -3.038 & 0.939 & 1.736 & -0.004 & 0 \\
			CrZrVAl & 6.430 & Half-metal & ----- & 0.366 & ----- & ----- & -3.305 & 0.611 & 2.217 & -0.010 & 0 \\
			CrHfVAl & 6.357 & Half-metal & ----- & 0.323 & ----- & ----- & -3.153 & 0.480 & 2.136 & -0.004 & 0 \\
			\hline
			%\midrule
			CrTiNbAl & 6.293 & SFM I & 0.115 & 1.244 & 0.115 & 0.210 & -2.717 & 1.273 & 0.815 & 0.023 & 0 \\
			CrZrNbAl & 6.527 & Half-metal & 0.110 & ----- & ----- & ----- & -2.874 & 0.905 & 1.144 & 0.023 & 0 \\
			CrHfNbAl & 6.450 & Half-metal & 0.040 & ----- & ----- & ----- & -2.593 & 0.611 & 1.017 & 0.028 & -0.718 \\
			\hline
			%\midrule
			CrTiTaAl & 6.260 & SFM I & 0.114 & 1.221 & 0.114 & 0.172 & -2.516 & 1.247 & 0.707 & 0.022 & 0 \\
			CrZrTaAl & 6.487 & Half-metal & 0.147 & ----- & ----- & ----- & -2.627 & 0.890 & 0.980 & 0.024 & -0.081 \\
			CrHfTaAl & 6.413 & Metal & ----- & ----- & ----- & ----- & -2.308 & 0.558 & 0.843 & 0.027 & -1.026 \\
			\specialrule{1pt}{0.5pt}{0.5pt}
			%\hline
			%\bottomrule % <-- Bottomrule here
			
			%\hline
			%\midrule
			CrAlVTi & 6.086 & Metal & ----- & ----- & ----- & ----- & -1.270 & 0.009 & 1.639 & -0.322 & 0.341 \\
			CrAlVZr & 6.375 & SFM I & 0.086 & 0.366 & 0.086 & 0.129 & -2.347 & 0.026 & 2.279 & -0.168 & 0 \\
			CrAlVHf & 6.301 & SFM II & 0.216 & 0.302 & 0.130 & 0.151 & -2.139 & 0.018 & 2.135 & -0.144 & 0 \\
			\hline
			%\midrule
			CrAlNbTi & 6.246 & SSC & 0.114 & 0.114 & 0.114 & 0 & 0 & 0 & 0 & 0 & 0 \\
			CrAlNbZr & 6.449 & SSC & 0.109 & 0.109 & 0.109 & 0 & 0.001 & 0 & -0.001 & 0 & 0 \\
			CrAlNbHf & 6.392 & SSC & 0.120 & 0.120 & 0.120 & 0 & 0 & 0 & 0 & 0 & 0 \\
			\hline
			%\midrule
			CrAlTaTi & 6.224 & SSC & 0.075 & 0.075 & 0.075 & 0 & -0.001 & 0 & 0 & 0 & 0 \\
			CrAlTaZr & 6.424 & SSC & 0.080 & 0.080 & 0.080 & 0 & 0 & 0 & 0 & 0 & 0 \\
			CrAlTaHf & 6.368 & SSC & 0.099 & 0.099 & 0.099 & 0 & -0.001 & 0 & 0.001 & 0 & 0 \\
			\specialrule{1pt}{0.5pt}{0.5pt}
			%\hline
			%\bottomrule % <-- Bottomrule here
			
		\end{tabular}
	\end{center}
\end{table*}

Figure~\ref{fig:struct} shows the crystal structure of a prototypical EQHA of spacegroup $\text{F}\bar{4}3m$ (216) with Wyckoff positions $a$, $b$, $c$, and $d$ occupied. For four distinct atomic species X, X’ , Y, and Z arranged across the four sites, 4! unique configurations are possible. We can further reduce this by noticing that cyclic permutations of the Wyckoff positions along the diagonal must form an equivalence relation between permutations, resulting in six equivalence classes. Lastly, half the remaining arrangements are related by inversion symmetry. Therefore, there are only three unique configurations of atomic species along the [111] body diagonal in the cubic Bravais lattice \cite{Stephen2019PRB}. To consistently label the various compounds we use the naming convention where XX'YZ corresponds to the 1st, 3rd, 2nd, and 4th site as shown in Fig.~\ref{fig:struct}. This notation is distinct from that used by Galanakis {\it et al.} in Ref.~\onlinecite{Galanakis2016}.  Each of the three unique arrangements along the [111] diagonal were studied for all combinations of Cr and Al with Group IVB (M = Ti, Zr, Hf) and Group VB elements (N = V, Nb, Ta), for a total of 27 unique quatenary Heusler compounds.

To classify the electronic phase of each compound we examine the relative energies of each spin population at valence and conduction band edges. The six distinct classifying phases are illustrated in Fig.~\ref{fig:phase}. A half-metal (Fig.~\ref{fig:phase}a)  is defined as any compound that is gapped in one spin channel and metallic for the opposite spin, yielding mobile spin polarized charge carriers. The formation of a zero-energy gap in a given spin channel can appear in two different ways. A Type I SGS (Fig.~\ref{fig:phase}b) has the valence band maximum of one spin channel coincide with its conduction band minimum, whereas a Type II SGS (Fig.~\ref{fig:phase}c) has one spin channel's valence band maximum coincide with the conduction band minimum of the opposite spin channel. In a fully gapped system, we define three unique cases. 
In a Type I SFM (Fig.~\ref{fig:phase}e) the valence band maximum and the conduction band minimum belong to the same spin channel. A Type II SFM (Fig.~\ref{fig:phase}f), on the other hand, has the valence band maximum and the conduction band minimum belong to different spin channels. Finally a symmetric semiconductor (SSC) (Fig.~\ref{fig:phase}d) exhibits no exchange splitting, or spin polarization at the Fermi level. In MTJ's employing magnetic insulators the properties of the occupied states are not believed to strongly influence the tunneling, rendering the distinction between Type I and II SFM's moot, but in situations where gating voltages are necessary, the two classes give rise to physically different situations.

The key to predicting strongly polarizing spin filters is the exchange splitting $2 \Delta E_{ex}$. Here, we define it as the energy difference between the conduction band minimum in each spin channel. Additionally, the overall size of the spin-dependent gaps, total band gap and effective masses are important in designing an effective spin filter.\cite{MullerFowNord} Intrinsic semiconductors with smaller band gaps typically have higher conductance, which is ideal for the advancement of spintronic devices\cite{microelec} while the differences in electron effective mass are rarely taken into account.\cite{Lukashev}

\subsection{Cr\textit{N}\textit{M}Al Ordered Arrangement}
All of the Cr\textit{N}\textit{M}Al ordered Heusler compounds are predicted to be fully-compensated spin filter materials (Table \ref{tab:table1}), with only CrVZrAl, CrVHfAl, and CrNbTiAl being Type I SFM candidates. This set of compounds also includes two arrangements originally predicted by Galanakis {\it et al}.\cite{Galanakis2014} In agreement with previous theoretical studies of CrVTiAl, we find a large exchange splitting of $0.430$ eV.\cite{Stephen2019PRB} This is the largest exchange splitting of all the compounds studied. CrTaHfAl exhibits an exceptionally small band gap of $0.120$ eV, which is half its exchange splitting and nearly 7 and 10 times smaller than CrVTiAl and EuO,\cite{MullerFowNord} respectively.
For this arrangement, the opening of the band gap is facilitated by the fully-compensated ferrimagnetic order stabilized over a tripartite lattice
formed by the Cr, Group VB, and Group IVB element sites. The Cr and Group VB atoms form a simple cubic cell in the quaternary Heusler structure, with the Group IVB atom or Al occupying
the body-center site. Upon substituting heavier Group VB elements, the size of the band gap in each spin channel decreases. As proposed in Ref. \onlinecite{Stephen2019PRB} for CrVTiAl, the dominant antiferromagnetic (AFM) interactions occur between the atoms on the simple cubic lattice between Cr and its next nearest neighbor. Therefore, as progressively heavier Group VB elements exhibiting weaker on-site correlation strengths are substituted in, the local magnetic moments and band gaps concomitantly reduce. The effect of introducing heavier Group IVB in the body-center site is more subtle and does not effect the magnetic moments or band gaps monotonically. 
\subsection{Cr\textit{M}\textit{N}Al Ordered Arrangement}
The second arrangement (Cr\textit{M}\textit{N}Al in Table \ref{tab:table1}) yields a mixture of half-metals and SGS/SFM candidate materials. CrZrVAl and CrHfVAl are found to be on the critical boundary between half-metal and SGS/SFM phases. Only a slight perturbation in the exchange interactions is necessary to facilitate a phase transition from half-metal to a Type I SGS or Type I SFM. The mixture of phases obtained for this arrangement, illustrates the competition between comparable AFM and ferromagnetic (FM) correlations. That is, since the position of the Group IVB and VB atoms are swapped with respect to the preceding case, the weaker correlations of the Group IVB elements reduce the AFM interactions, while the more locally correlated Group VB elements develop a sizable FM double exchange interaction on their respective sub-lattice. This competition then drives the overall weaker ordering moments leading to SGS/half-metallic behaviors. CrTiNbAl and CrTiTaAl are interesting exceptions with Ti having an surprisingly large moment in these compounds leading to a gap being maintained in both spin channels. Both have small band gaps of 0.115 eV and 0.114 eV with exchange splitting of 0.210 eV and 0.172 eV, respectively.

\begin{figure}[t]
\includegraphics[width=0.97\columnwidth]{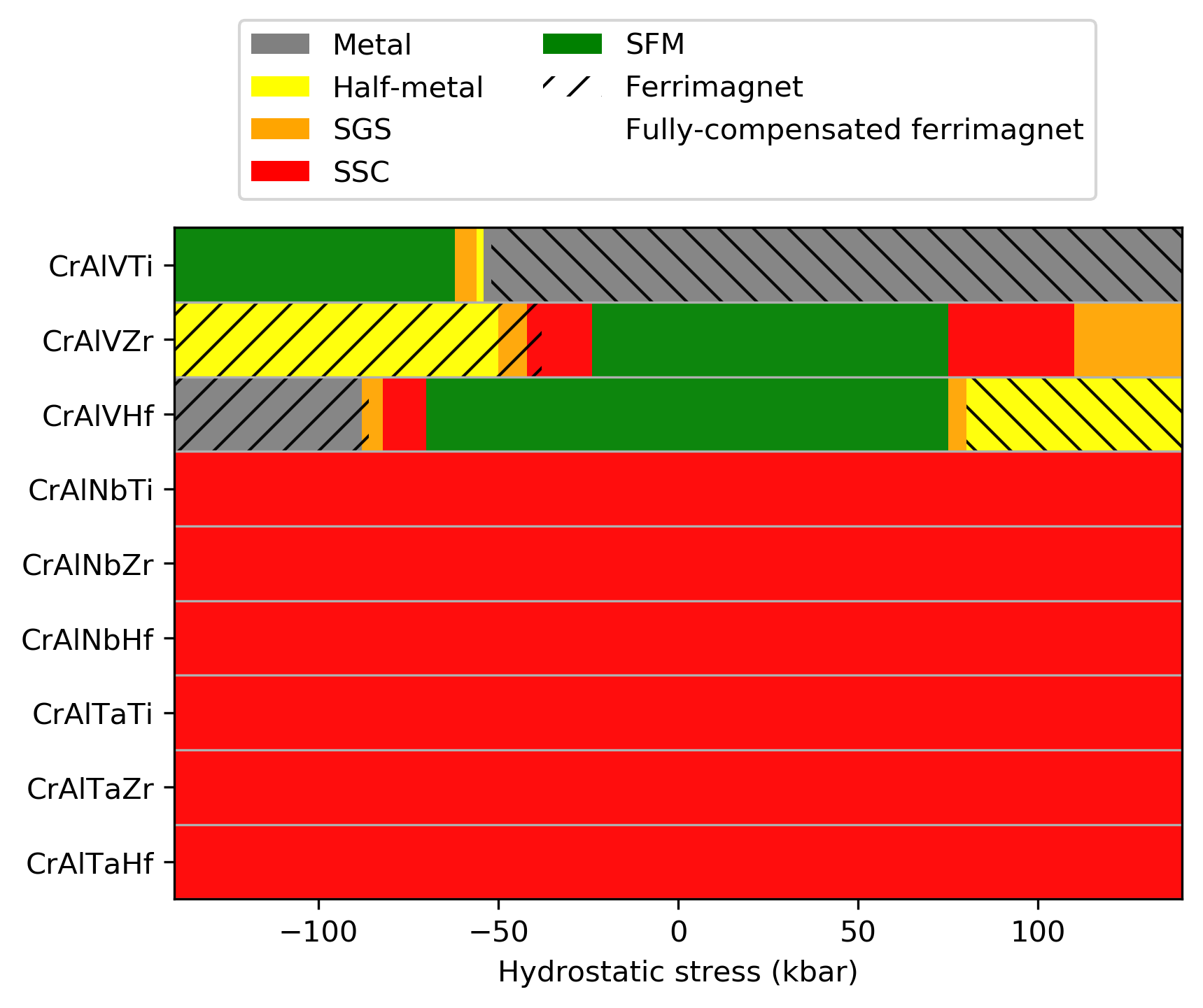}
\caption{(color online) Phase diagram of predicted electronic phases as a function hydrostatic stress. The electronic and magnetic states are indicated by color and overlaid pattern, respectively.}
\label{fig:phasediagram}
\end{figure}

\subsection{CrAl\textit{N}\textit{M} Ordered Arrangement}
Finally, most EQHA's in the CrAl\textit{N}\textit{M} arrangement in Table \ref{tab:table1} are symmetric semiconductors with small band gaps, and thus do not exhibit any advantageous spin filter properties. For this configuration, AFM correlations have all but been quenched completely by swapping Cr in to the body-centered position, with the Group IVB and VB elements occupying the simple cubic sites. Since Al has no $d$ electrons and is inert magnetically, Cr can only couple to its self ferromagnetically. Thus, the lack of an AFM interaction with Cr leads to nonmagnetic insulating systems or FM metals. Interestingly, compounds containing V with Zr or Hf are fully-compensated ferrimagnetic SFM's and the compound containing Ti has a small uncompensated moment of 0.341$\mu_{B}$ and a metallic phase. In this family, Vanadium is the element with the second largest magnetic moment compared to Cr. Therefore, in the absence of Cr on the simple cubic lattice V drives the AFM correlations with its next nearest neighbors. This is clearly illustrated by the fact that the magnetic moments on the Cr and the Group IVB atomic sites are collinear with respect to the local moment on the V site. In the case of CrVTiAl, however, the AFM correlations are not strong enough to fully gap the electronic states.

Overall, we find a delicate balance between the AFM and FM correlations, which drives the characteristic differences between the SFM, SGS, SSC, and metallic phases. In the SFM phase, the two strongest magnetic elements sit on the same simple cubic lattice, enabling strong AFM coupling. In the SGS and metallic phases, the magnetic elements are moved to the body-center position, weakening the AFM correlations. Moreover, as the stronger magnetic atoms are placed in line with the uncorrelated inert Al, FM double-exchange interactions are enhanced. The interplay of these two opposing interactions is reflected in the electronic structure through the gradual closing of the various spin-dependent band gaps. Additionally, as elements are substituted for heavier equivalent atoms, correlation effects are reduced due to the delocalization of the $d$-orbital electrons. In select cases, this leads to possible hybridization driven magnetism.  \cite{williams1981covalent}

Finally, CrVZrAl, CrVHfAl, CrTiNbAl, and CrTiTaAl are found to be spin filters in 2 out of the 3 configurations. Thus making these combinations partially immune to disorder as compared to CrVTiAl. Therefore, even though they have a smaller exchange splitting they are better candidates to successfully realize a working SFM in thin films. The small band gaps of CrTiNbAl and CrTiTaAl should make them especially attractive.

\section{Effect of Hydrostatic Stress on Electronic Structure}\label{sec:stress}
Since the CrVAl\textit{M} subfamily of compounds, along with their possible arrangements, display a diverse array of electronic phases, we gauge their stability and tunablity to external perturbations by applying hydrostatic compressive and tensile stress. Compressive stress was applied in increments of 5 kbar, while tensile stress was applied in increments of 2 kbar to mitigate numerical instabilities. The cell shape and volume along with the internal atomic degrees of freedom were relaxed for each step in pressure. Here, negative (positive) hydrostatic stress corresponds to tensile (compressive) stress. The applied stresses are readily experimentally accessible by induced substrate lattice mismatch during epitaxial growth or through the use of diamond anvil cells.\cite{Suwardi} 

\begin{figure}[ht!]
\begin{subfigure}{.5\textwidth}
  \centering
  \includegraphics[width=0.97\columnwidth]{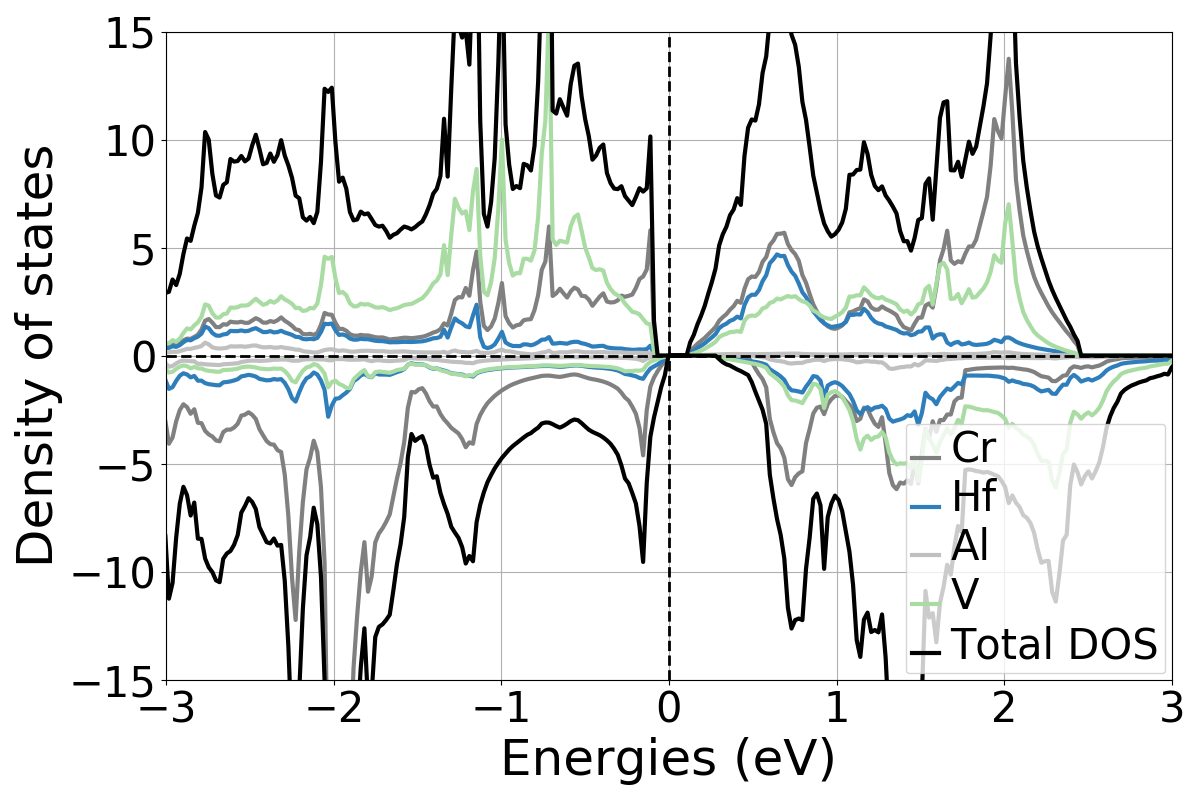}
  \caption{}
  \label{fig:dos1}
\end{subfigure}
\begin{subfigure}{.5\textwidth}
  \centering
  \includegraphics[width=.97\columnwidth]{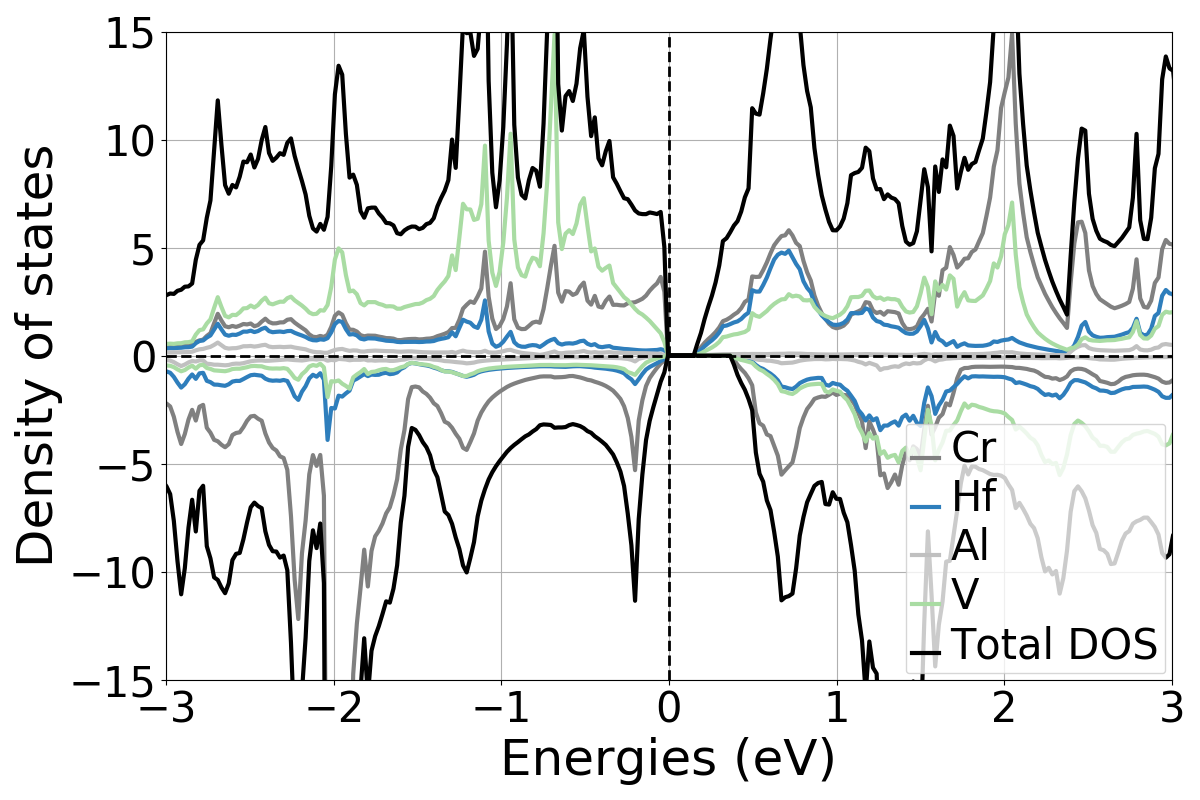}
  \caption{}
  \label{fig:dos2}
\end{subfigure}
\begin{subfigure}{.5\textwidth}
  \centering
  \includegraphics[width=.97\columnwidth]{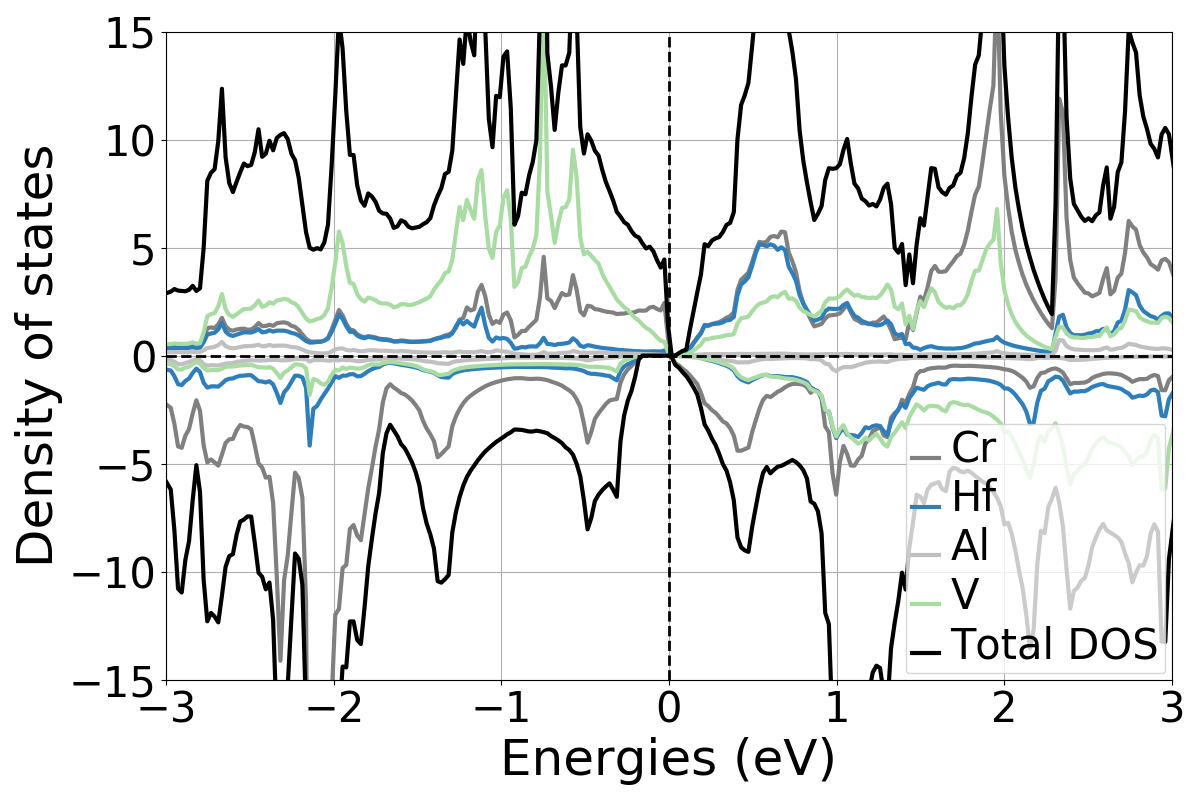}
  \caption{}
  \label{fig:dos3}
\end{subfigure}
\caption{(color online) Element-projected densities of states for CrAlVHf under (a) 0.0 kbar, (b) 30 kbar, and (c) 86 kbar of applied tensile stress.}
\label{fig:doss}
\end{figure}

Figure \ref{fig:phasediagram} shows a phase diagram of the all the CrAlV\textit{M} subfamily compounds as a function of pressure. Strikingly, the compounds containing V display the most varied behavior. The electronic phase is quite tunable with applied stress, exhibiting all electronic phases over a mild stress range. Starting at zero stress the majority spin band gaps in the SFM ground state of CrAlVZr and CrAlVHf quickly decrease with stress, precipitating a phase transition to an SGS state upon their closure. Subsequently for larger compressive (tensile) stress, a ferrimagnetic half-metal phase appears in CrAlVZr while CrAlVHf becomes a ferrimagnetic normal metal. CrAlVTi, on the other hand is a metal in the ground state and by applying tensile stress a transition opening a gap with small exchange splitting is induced. This suggests that systems synthesized with slight atomic site disorder may be tuned by a small applied stress into a SFM. Moreover, stress may be used to switch the electronic state from SFM to metal or half-metal allowing for greater control over spin polarized currents. The main effect of the pressure is to influence the magnitude of the magnetic moment of each atom. This behavior is most clearly seen for CrAlVTi. The phase change from metal to SFM in CrAlVTi can be attributed to the strengthening of the antiferromagnetic correlations of V and Ti due to the increase of their magnetic moments. 

Figure~\ref{fig:doss} compares the atomic site projected density of states (DOS) for CrAlVHf under $0$, $30$ and $86$ kbar of applied tensile stress. The various colored lines denote the DOS for the different atomic species. By comparing Figures~\ref{fig:dos1} and \ref{fig:dos2} the effect of $30$ kbar of tensile stress is readily seen. The relative energies of the Cr and V character states in the majority-spin channel are shifted to slightly higher energies, marking an increase in the exchange splitting. Figures~\ref{fig:dos2} and \ref{fig:dos3} show the DOS for $30$ kbar and $86$ kbar of stress, respectively. Here a clear switch in the electronic phase is seen with pressure. For increasing tensile stress the band gap energy shrinks in both spin channels converting CrAlVHf into a Type II SGS and then for larger stress a metallic phase appears. Coincident with the transition from type II SGS to metal, a magnetic transition occurs at approximately 90 kbar and the cell is no longer fully compensated. 

\section{Concluding Remarks}\label{sec:conclusion}
By calculating the electronic and magnetic properties of 27 new EQHA's we identify several new possible spin filter materials, along with other SGS and half-metal phases. Importantly, CrVZrAl, CrVHfAl, CrTiNbAl, and CrTiTaAl are partly immune to atomic site disorder important to synthesizing single crystals and thin films. Overall, we find the ground state electronic phases intimately connected to the competition between AFM and FM correlations. This should serve as an accurate descriptor to search for other new SFM.

%% Acknowledgments

\begin{acknowledgments}
The authors thank Dr. Gregory Stephen and Prof. Don Heiman for many fruitful discussions.
The work at Northeastern University was supported by the U.S. DOE, Office of Science, Basic Energy Sciences grant number DE-FG02-07ER46352 (core research) and benefited from Northeastern University’s Advanced Scientific Computation Center, the National Energy Research Scientific Computing Center supercomputing center (DOE grant number DEAC02-05CH11231), and support (testing the efficacy of new functionals in diversely bonded materials) from the DOE Energy Frontier Research Centers: Center for the Computational Design of Functional Layered Materials (DE-SC0012575). The work at Los Alamos National Laboratory was supported by the U.S. DOE NNSA under Contract No. 89233218CNA000001 through the LDRD program and by the Center for Integrated Nanotechnologies, a DOE BES user facility.
\end{acknowledgments}

\bibliography{Heuslers_Refs}

\end{document}